\documentclass[a4paper,12pt]{article} 

\usepackage{geometry,epsfig,subfigure}
\usepackage{amsmath}
\usepackage{amssymb}
\usepackage{graphics,epsfig,mydef}

\usepackage[square,sort&compress,comma,numbers]{natbib}

\usepackage[usenames]{color}
\usepackage{fancyhdr,url}   
\usepackage{setspace}  
\usepackage{multirow}   
\usepackage{rotating}    
\usepackage{colortbl}     
\usepackage{relsize} 
\usepackage{booktabs}
\usepackage{cancel}
\usepackage{enumerate}
\numberwithin{equation}{section}
\usepackage{soul} 
\usepackage{xcolor} 
\usepackage{pdflscape}
\usepackage{hyperref}
\usepackage[capitalise]{cleveref}
\usepackage{lineno}
\usepackage{graphicx}
\usepackage{etoolbox}
\usepackage{epstopdf}

\definecolor{darkblue}{rgb}{0,0,0.8}
\definecolor{darkgreen}{rgb}{0,0.5,0}

\long\def\symbolfootnote[#1]#2{\begingroup \def\thefootnote{\fnsymbol{footnote}}\footnote[#1]{#2} \endgroup} 

\usepackage{amsthm}

\newcommand{\HRule}{\rule{0.9\linewidth}{0.2mm}}

\usepackage{nomencl}
\makenomenclature

\usepackage{etoolbox}
\renewcommand\nomgroup[1]{%
  \item[
  \ifstrequal{#1}{A}{\textit{Symbols}}{%
  \ifstrequal{#1}{B}{\textit{Greek symbols}}{%
  \ifstrequal{#1}{C}{\textit{Subscripts}}{}}}%
]}

\begin{document}
\renewcommand*{\thepage}{\arabic{page}}

\setstretch{1.3}

\begin{center}
\large
\textbf{A theoretical framework for comprehensive modeling of steadily fed evaporating droplets and the validity of common assumptions\\}

\normalsize
\vspace{0.2cm}
Yigit Akkus$^{a}$, Barbaros \c{C}etin$^{b}$, Zafer Dursunkaya$^{c}\symbolfootnote[1]{e-mail: \texttt{refaz@metu.edu.tr}}\!$\\
\smaller
\vspace{0.2cm}
$^a$ASELSAN Inc., 06200 Yenimahalle, Ankara, Turkey\\
$^b$Mechanical Engineering Department, \.I.D. Bilkent University, 06800  \c Cankaya, Ankara, Turkey\\
$^c$Department of Mechanical Engineering, Middle East Technical University, 06800 \c Cankaya, Ankara, Turkey\\
\vspace{0.2cm}
\end{center}

\begin{center} \noindent \HRule \\ \end{center}
\vspace{-0.6cm}
\begin{abstract}

\noindent 

%
%
%
%
%
%
%

A theoretical framework is established to model the evaporation from continuously fed droplets, promising tools in the thermal management of high heat flux electronics. Using the framework, a comprehensive model is developed for a hemispherical water droplet resting on a heated flat substrate incorporating all of the relevant transport mechanisms: buoyant and thermocapillary convection inside the droplet and diffusive and convective transport of vapor in the gas domain. At the interface, mass, momentum, and thermal coupling of the phases are also made accounting for all pertinent physical aspects including several rarely considered interfacial phenomena such as Stefan flow of gas and the radiative heat transfer from interface to the surroundings. The model developed utilizes temperature dependent properties in both phases including the density and accounts for all relevant physics including Marangoni flow, which makes the model unprecedented. Moreover, utilizing this comprehensive model, a nonmonotonic interfacial temperature distribution with double temperature dips is discovered for a hemispherical droplet having internal convection due to buoyancy in the case of high substrate temperature. Proposed framework is also employed to construct several simplified models adopting common assumptions of droplet evaporation and the computational performance of these models, thereby the validity of commonly applied simplifying assumptions, are assessed. Benchmark simulations reveal that omission of gas flow, \textit{i.e.} neglecting convective transport in gas phase, results in the underestimation of evaporation rates by 23--54\%. When gas flow is considered but the effect of buoyancy is modeled using Boussinesq approximation instead of assigning temperature dependent density throughout the gas domain, evaporation rate can be underestimated by up to 16\%. Deviation of simplified models tends to increase with increasing substrate temperature. Moreover,    presence of Marangoni flow leads to larger errors in the evaporation rate prediction of simplified models.

\vspace{0.2cm}

\end{abstract}
\vspace{-0.6cm}
\begin{center} \noindent \HRule \\ \end{center}

\pagebreak


\nomenclature[C]{$c$}{capillary}
\nomenclature[C]{$d$}{disjoining}
\nomenclature[C]{$l$}{liquid}
\nomenclature[C]{$lv$}{liquid-vapor}
\nomenclature[C]{$v$}{vapor}
\nomenclature[C]{$w$}{wall}
\nomenclature[C]{$s$}{first derivative with respect to $s$}
\nomenclature[C]{$ss$}{second derivative with respect to $s$}
\nomenclature[C]{$sss$}{third derivative with respect to $s$}

\section{Introduction}
\label{sec:intro}

Droplet evaporation is a ubiquitous phenomenon observed in various natural processes such as human perspiration and industrial  applications such as DNA mapping, inkjet printing, biosensing, and surface coating \cite{smalyukh2006,lim2009evaporation,ebrahimi2013,wu2014}. In recent years, utilization of droplet evaporation in cooling applications has been of interest due to its high heat removal capability associated with the latent heat of vaporization during phase change. While some studies propose the use of drying droplets such as spray cooling \cite{bar2006,liang2017}, others suggest the utilization of continuously fed, constant shape droplets \cite{kokalj2010,macdonald2017b} similar to sweat-droplets on mammals' skin. In the absence of feeding, droplet evaporation is a transient process because of the deforming droplet surface. However, in the case of steadily fed droplets, liquid-gas interface preserves its shape and the problem becomes that of a steady state configuration.

Regardless of being transient or steady state, droplet evaporation is a complex problem because of the presence of various energy transport mechanisms in two different phases together with their coupling at the liquid-gas interface where heat and mass transfer take place simultaneously. Inside the droplet, energy is transferred from the substrate to the interface \textit{via} conduction and convection. The latter was rarely considered before 2000's \cite{duh1989,lozinski1993,shih1996} but subsequent studies \cite{ruiz2002,girard2006,hu2005,xu2007,ristenpart2007,kaneda2008,xu2009,yoshitake2010,lu2011,zhang2014temperature,barash2015,bouchenna2017,josyula2018,wang2018,akkus2019iterative} focused on the convective heat transport inside the droplets, which is triggered by two mechanisms; buoyancy and thermocapillarity \cite{ruiz2002}. Thermocapillary convection was reported to dominate the buoyancy and dictate the flow pattern inside an evaporating droplet \cite{lu2011,bouchenna2017}. Outside the droplet, energy is transferred from the droplet surface to the ambient \textit{via} diffusion of vapor, convection of gas, and conduction in gas phase. In the absence of a forced flow of surrounding gas, natural convection is responsible for the convective heat transfer in the gas phase. While the majority of modeling attempts did not account for the effect of buoyancy in gas phase, several experiments \cite{kelly2011,sobac2012pre,carle2013} demonstrated that diffusion-controlled evaporation models considerably underestimate the evaporation rate. In addition, this issue was also confirmed by empirical \cite{carle2016} and numerical \cite{saada2010,chen2017} models. 

The proper coupling of the phases at the interface is essential in building a successful computational model; therefore, mass,    force, and energy balances should be properly imposed at the liquid-vapor interface. The mass balance results in the discontinuity of normal velocities of liquid and vapor due to the density difference of the phases. Moreover, interfacial gas velocity is dictated by the Stefan flow of air to counteract the diffusion of air towards the interface, through which air cannot penetrate due to its insolubility in liquid. Thermocapillarity produces a surface force in the tangential direction and the curved surface creates a Laplace pressure jump; therefore, tangential and normal force balances need to be constructed accordingly. Energy transferred to the liquid interface is conveyed to the gas phase/surroundings by evaporative heat transfer, conduction to the gas phase and radiation to the surroundings. While evaporation is always responsible for the majority of the heat transfer, the contribution of the others increase when the temperatures of the interface and substrate increase. 

An accurate estimation of evaporative mass flux is vital since it is the major energy transfer mechanism at the interface. Although both diffusive and convective components contribute to the total evaporative flux, majority of the previous studies \cite{hu2002evap,ristenpart2007,xu2009,yoshitake2010,zhang2014temperature,barash2015,bouchenna2017,josyula2018,wang2018} neglected the effect of convection. Many of them \cite{xu2009,yoshitake2010,zhang2014temperature,barash2015,wang2018} implemented the semi-empirical correlation of Hu and Larson \cite{hu2002evap}, which was built on the well-known studies of Deegan \textit{et al.} \cite{deegan1997,deegan2000}. In the derivation of this correlation, the sole transport mechanism considered in the gas phase was the diffusion of vapor. Moreover, the droplet surface was assumed to be isothermal and the correlation was obtained for contact angles less than 90$^\circ$. Another vapor diffusion based correlation was suggested by Popov \cite{popov2005}, which accounted for the nonuniformity of evaporative mass flux at the droplet surface and valid for all contact angles. This correlation was also used in subsequent studies \cite{carle2016,josyula2018}. Estimation of convective mass transfer, on the other hand, requires the solution of flow in the gas phase. Among the studies carrying out the solution of gas flow \cite{saada2010,bouchenna2017,chen2017,pan2020} in droplet evaporation problems, only a few \cite{chen2017,pan2020} considered the convective transport in the calculation of evaporation rate. In addition, several studies \cite{ruiz2002,girard2006,girard2008,kaneda2008,akkus2017modeling,akkus2019iterative} implemented correlations based on analogies between heat and mass transfer instead of calculating a mass diffusion based solution of concentration field in the gas phase. Lastly, the kinetic theory of gases was also adopted in the prediction of the evaporation flux in the literature \cite{lu2011,strotos2008}.

The present study aims to develop a comprehensive model in order to estimate the evaporation rate from steadily fed evaporating droplets. In our previous models, we only considered the liquid flow and energy transfer inside the droplet without \cite{akkus2017modeling} and with the thermocapillarity effect \cite{akkus2019iterative}, and the evaporative mass flux at the interface was estimated based on an analogy between heat and mass transfer proposed by \cite{ruiz2002}. The current study takes a further step and includes the solution of mass, species, momentum and energy transfer in the gas mixture. The resultant concentration field of the vapor and flow field of the gas are utilized to estimate the evaporation rate from a droplet. The computational model developed is tested by considering the evaporation from a spherical (with 90$^{\circ}$ contact angle), sessile and continuously fed water droplet---by liquid injection---placed on a heated flat substrate, which was the same configuration in our previous work \cite{akkus2019iterative} and a previous experimental work \cite{macdonald2017} thereby enables a direct comparison with previous results.

The ultimate objective of the current work is to develop a general theoretical framework for the modeling steadily fed evaporating droplets by incorporating all relevant physical phenomena inside and outside the droplet as well as with the ones at the interface. Since the steadily fed droplets are proposed in the thermal management of high heat flux dissipating electronic components, the substrate temperature is expected to reach high values. Although Boussinesq approximation \cite{saada2010,carle2016,bouchenna2017,chen2017} is an alternative approach to solving full compressible Navier-Stokes equations to simplify  nonlinearity and improve the numerical convergence, strictly speaking, it is not suitable for air when the temperature difference between the heated surface and the far field is higher than $15\degC$ \cite[p. 14-15]{Ferziger02}. The computational model presented utilizes full compressible Navier-Stokes equations with temperature dependent thermophysical properties. Therefore, buoyancy effects in both phases together with the varying surface tension along the droplet surface (thermocapillarity) are simulated without any approximation, which makes the present work unprecedented. Moreover, in order to assess the validity of these common assumptions, the results of the proposed model is compared with several simplified models and correlations applying widely used assumptions in the literature .

The article is organized as follows: in \cref{sec:modeling}, modeling strategy is presented by providing detailed information about the governing equations and associated boundary conditions in both liquid and gas domains. In \cref{sec:scheme}, the iterative computational scheme is explained step by step. The resultant flow, temperature, and concentration fields for a certain test configuration are reported and elucidated with and without the presence of Marangoni convection for two different substrate temperatures in \cref{sec:results}. Moreover, the framework proposed is utilized to make a benchmark test for the models utilizing simplifying assumptions in droplet evaporation modeling to question the validity of these common assumptions. Finally, summary and conclusions are presented in \cref{sec:conclusion}.   

\section{Theoretical Modeling}
\label{sec:modeling}

\begin{figure}[h]
\includegraphics[scale=0.9]{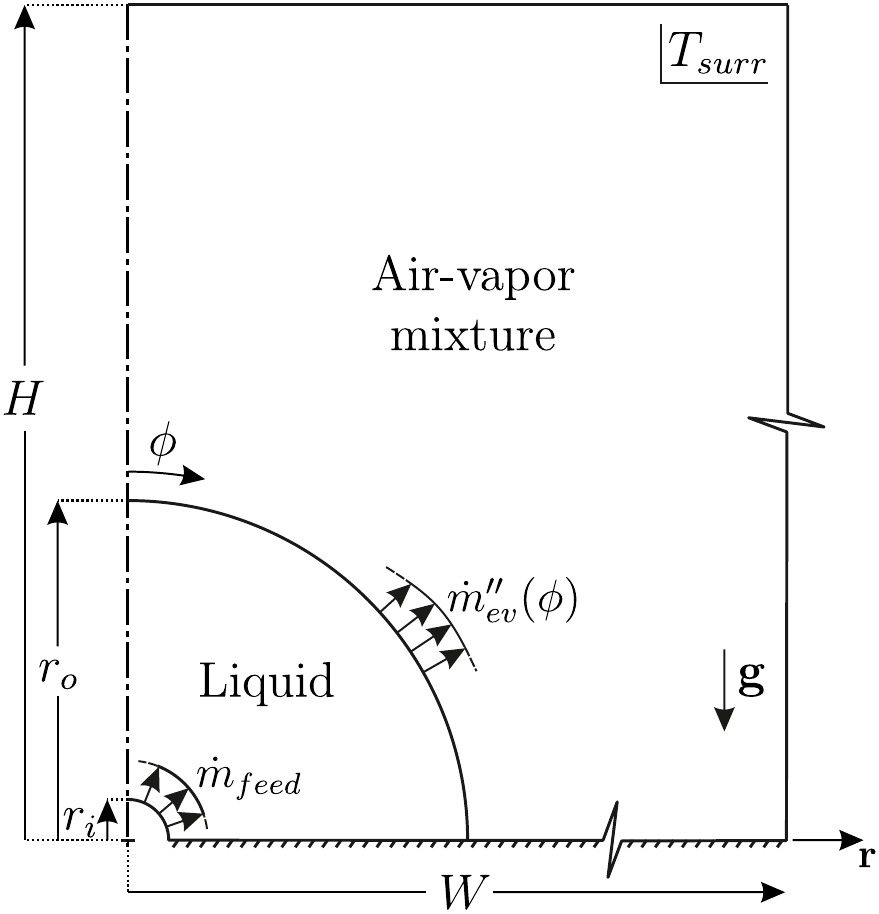}
\centering
\caption{Problem domain}
\label{fgr:domain}
\end{figure}

Steady evaporation from a steadily fed liquid droplet resting on a heated substrate to ambient air is modeled. As long as the capillary forces dominate the gravitational ones, the Bond number is smaller than unity and the droplet surface assumes a spherical shape. In our model, a hemispherical droplet is considered using a 2-D axisymmetric model as shown in \cref{fgr:domain}. A large air volume in the shape of a cylinder encloses the droplet and the flat substrate. Since steadily fed droplets are considered in the thermal management of electronic components, material of the substrate (heat sink) is likely to be a thermally highly conductive metal, which causes a nearly constant substrate temperature. Therefore, constant wall temperature is assigned to the substrate surface. Moreover, feeding liquid is assumed to be in thermal equilibrium with the substrate.

\newpage
\subsection{Coupling of the Phases}
\label{coupling}

\subsubsection{Mass balance at the interface}
\label{mass}

In the liquid phase, normal component of the interfacial velocity  determines the evaporating mass flux ($\dot{m}''_{ev}$):

\begin{equation} \label{eqn:mass_balance}
\mathbf{u_{\ell}} \cdot \mathbf{n}=\frac{\dot{m}''_{ev}}{\rho} \, ,
\end{equation}

\noindent where $\mathbf{n}$ and $\rho$ are unit normal vector and density of the liquid at the interface, respectively, and the subscript $\ell$ designates the liquid phase. When the value of evaporating mass flux is provided, \cref{eqn:mass_balance} can be used to determine the normal velocity of the liquid phase at the interface. However, the same equation cannot be used to evaluate the interfacial gas velocity due to the presence of additional air flow from the interface to the surrounding gas. This air flow, also known as Stefan flow, is determined by equating the diffusive air transport towards the interface and the convective air transport from the interface: 

\begin{equation} \label{eqn:stefan}
D  (\nabla \cdot  \mathbf{n})c_{air} = (\mathbf{u}_g \cdot \mathbf{n}) c_{air} \, ,
\end{equation}

\noindent where $c_{air}$ is the molar concentration of air and $D$ is the binary diffusion coefficient; the subscript $g$ designates the gas phase. \Cref{eqn:stefan} is used to estimate the normal component of the gas velocity, \textit{i.e.} $u_n=(D/c_{air})(\partial c_{air}/ \partial n)$, in the current study. Tangential gas velocity, on the other hand, is equal to the tangential liquid velocity \cite{prosperetti1979,carle2016}, which is determined from the solution of the governing equations in liquid domain. The primary factor determining the tangential liquid velocity is the tangential force balance at the interface. 
  
\subsubsection{Force balance at the interface}
\label{momentum}
At the interface between the liquid and gas phases, normal and tangential stress balances are given, respectively, as follows:

\begin{subequations}
\begin{equation} \label{eqn:force_normal}
\mathbf{n} \cdot {\bar{\bar{\tau}}}_g \cdot \mathbf{n} -\mathbf{n} \cdot {\bar{\bar{\tau}}}_{\ell} \cdot \mathbf{t} = \gamma \nabla \cdot \mathbf{n} \, ,
\end{equation}
\begin{equation} \label{eqn:force_tangent}
\mathbf{n} \cdot {\bar{\bar{\tau}}}_g \cdot \mathbf{n} -\mathbf{n} \cdot {\bar{\bar{\tau}}}_{\ell} \cdot \mathbf{t} = \nabla\gamma \cdot \mathbf{t} \, ,
\end{equation}
\end{subequations}

\noindent where $\bar{\bar{\tau}}$ is the deviatoric stress tensor defined as \mbox{$\mu ( \partial u_{i} / \partial x_{j} + \partial u_{j} / \partial x_{i} )$}, $\gamma$ is the surface tension, and $\mathbf{t}$ is the unit vector in tangential direction. The normal stress balance expresses the interplay of normal forces associated with the pressures (including the Laplace pressure due to curved interface) and normal velocities in the two phases. Normal stress balance, however, is not implemented in the model proposed by the current study for two reasons: the normal velocities in both phases are determined from the mass balance relations, and accurate pressure transition between phases is not critical due to the fact that liquid and gas domains are solved separately. Application of tangential stress balance, on the other hand, is essential since the non-uniform distribution of the interfacial temperature initiates thermocapillary (or Marangoni) convection. The shear stress induced by the gas phase is assumed to be much smaller than that of the liquid phase. The validity of this assumption is confirmed by an \textit{a posteriori} analysis of the results. Therefore, the effect of gas shear on the interface force balance is neglected.  

\subsubsection{Energy balance at the interface}
\label{energy}
Energy transfer from liquid interface to the ambient occurs \textit{via} three mechanisms: (i) evaporative heat (mass) transfer, (ii) conduction to the gas phase, and (iii) radiation to the surroundings. The resultant energy balance at the interface is provided below: 

\begin{equation} \label{eqn:energy_balance}
\mathbf{n} \cdot (-k_l \nabla l) = \dot{m}''_{ev}h_{fg}-\mathbf{n} \cdot (-k_g \nabla T_g) 
+\sigma \epsilon (T_s^4-T_{surr}^4) \, ,
\end{equation}

\noindent where $k$, $h_{fg}$, $\sigma$, and $\epsilon$ are thermal conductivity, latent heat of vaporization, Stefan-Boltzmann constant, and emissivity of the liquid surface; $T_s$ and $T_{surr}$ are the temperatures of the interface and surroundings, respectively. Evaporative heat transfer dominates the others due to the high latent of vaporization. Therefore, estimation of evaporative mass flux ($\dot{m}''_{ev}$) has paramount effect on the temperature and flow fields in both domains. Evaporative mass flux is determined by the summation of diffusive and convective transfer of the vapor from the interface:

\begin{equation} \label{eqn:evap_flux}
\dot{m}''_{ev}= -D  (\nabla \cdot  \mathbf{n})c_v  + (\mathbf{u_g} \cdot \mathbf{n}) c_v \, ,
\end{equation}

\noindent where $c_{v}$ is the molar concentration of vapor. Therefore, estimation of evaporation flux requires the solution of the concentration field of vapor in the gas domain. 

\subsection{Governing Equations and Boundary Conditions}

Steady forms of the conservation equations for mass, linear momentum, and energy in both liquid and gas phases together with species conservation equation for vapor transport in the gas domain are solved. Governing equations are summarized below:

\begin{subequations}
\begin{equation} \label{eqn:vector-cont}
\nabla\cdot(\rho\mathbf{u}) = 0
\end{equation}
\begin{equation} \label{eqn:vector-mom}
\rho(\mathbf{u}\cdot\nabla)\mathbf{u} = -\nabla p + \nabla \cdot \bar{\bar{\tau}} + \rho\mathbf{g}
\end{equation}
\begin{equation} \label{eqn:vector-en}
\rho c_p\mathbf{u}\cdot\nabla T = \nabla \cdot (k\nabla T) + \bar{\bar{\tau}}:\nabla \mathbf{u}
\end{equation}
\begin{equation} \label{eqn:vector-species}
\mathbf{u}\cdot\nabla c_v = \nabla \cdot (D\nabla c_v)
\end{equation}
\end{subequations}

\noindent where $\mathbf{g}$ is the gravitational acceleration; $\mu$ and $c_p$ are dynamic viscosity and specific heat, respectively. All fluid properties, including the density, are defined as temperature dependent and the values are taken from the material library of COMSOL Multi-physics software.

Symmetry boundary condition is applied along the center line of the droplet and the surrounding air volume. No slip and constant temperature boundary conditions are used on the substrate. The velocity of the feeding liquid ($\mathbf{\bar u_{in}}$) is assumed to be uniform and the temperature of the feeding liquid is equal to wall temperature ($T_w$). At the liquid-gas interface mass, tangential force, and energy balances are secured as explained in Section~2.1. Boundary conditions for the liquid domain are summarized below:  

\begin{subequations}
\begin{equation} \label{eqn:liq-b1}
\partial_\phi \mathbf{u} =0 \hskip 1 pt ; \hskip 3 pt  \partial_\phi T=0 \hskip 5 pt {\rm at} \hskip 5 pt \phi=0
\end{equation}
\begin{equation} \label{eqn:liq-b2}
\mathbf{u}=0 \hskip 1 pt ; \hskip 3 pt T=T_w \hskip 5 pt {\rm at} \hskip 5 pt  \phi=\pi/2
\end{equation}
\begin{equation} \label{eqn:liq-b3}
\mathbf{u}=\mathbf{\bar u_{in}} \hskip 1 pt ; \hskip 3 pt T=T_w \hskip 5 pt {\rm at} \hskip 5 pt  r=r_i
\end{equation}
\begin{eqnarray} \label{eqn:liq-b4}
&\mathbf{u} \cdot \mathbf{n}=\dot{m}''_{ev}/{\rho} \hskip 1 pt , -\mathbf{n} \cdot \bar{\bar{\tau}} \cdot \mathbf{t}=\nabla\gamma \cdot \mathbf{t} \hskip 1 pt ; \nonumber \\
&\mathbf{n} \cdot (-k_l \nabla T) = \dot{m}''_{ev}h_{fg}-\mathbf{n} \cdot (-k_g \nabla T_g) 
+\sigma \epsilon (T^4-T_{surr}^4) \hskip 5 pt {\rm at} \hskip 5 pt  r=r_o
\end{eqnarray}
\end{subequations}

\noindent Evaporative mass flux ($\dot{m}''_{ev}$) requires the solution of concentration and flow fields in the gas domain. Therefore, evaporative mass flux together with the temperature distribution in the gas phase ($T_g$) are unknown \textit{a priori}. They are evaluated in an iterative algorithm whose details are given in Section~3. Temperature of the surroundings ($T_{surr}$) is assumed to be equal to the ambient gas temperature. 

In the gas phase, thermal and hydrodynamic boundary conditions identical to those of the liquid phase are utilized at the center line and substrate surface. In addition, no vapor penetration condition is used at these boundaries. At the outer boundaries, gas temperature, gas pressure, and vapor concentration are assumed to reach their ambient values. At the liquid-gas interface, the phases are considered to be at thermal equilibrium; therefore, surface temperature estimated from the solution of liquid domain ($T_s$) is assigned to the gas domain. A gas velocity distribution ($\mathbf{u_s}$) is assigned to the interface by combining its normal and tangential components. While normal component is calculated based on the Stefan flow (see \cref{eqn:stefan}), tangential one is taken from the solution of the velocity field in the liquid domain. Boundary conditions for the gas domain are summarized below: 

\begin{subequations}
\begin{equation} \label{eqn:gas-b1}
\partial_\phi \mathbf{u} =0 \hskip 1 pt ; \hskip 3 pt  \partial_\phi T=0 \hskip 1 pt ; \hskip 3 pt \partial_\phi c_v=0 \hskip 5 pt {\rm at} \hskip 5 pt \phi=0
\end{equation}
\begin{equation} \label{eqn:gas-b2}
\mathbf{u}=0 \hskip 1 pt ; \hskip 3 pt T=T_w \hskip 1 pt ; \hskip 3 pt \partial_\phi c_v=0 \hskip 5 pt {\rm at} \hskip 5 pt  \phi=\pi/2
\end{equation}
\begin{equation} \label{eqn:gas-b3}
p=p_{\infty} \hskip 1 pt ; \hskip 3 pt T=T_{\infty} \hskip 1 pt ; \hskip 3 pt c_v= \phi_{RH} c_{v,sat} \hskip 5 pt {\rm at} \hskip 3 pt  {\rm the} \hskip 3 pt  {\rm outer} \hskip 3 pt {\rm boundaries}
\end{equation}
\begin{equation} \label{eqn:gas-b4}
\mathbf{u}=\mathbf{u_s} \hskip 1 pt ; \hskip 3 pt T=T_s  \hskip 1 pt ; \hskip 3 pt c_v=c_{v,sat}  \hskip 5 pt {\rm at} \hskip 5 pt  r=r_o
\end{equation}
\end{subequations}

\noindent where $\phi_{RH}$ is the relative humidity in the far field and $c_{v,sat}$ is the saturation concentration of vapor at the corresponding temperature.

\section{Computational Scheme} 
\label{sec:scheme}
Primary challenge in the modeling of droplet evaporation is the coupling of a condensed phase with a gas phase. Transition between the phases should be delicately treated since the resultant flow, temperature, and concentration fields are shaped based on the dynamics at the interface. One way to mitigate the coupling problem can be the utilization of a proper two phase method, which deals with the two phases simultaneously, but this approach, undoubtedly, brings a higher computational cost. An interface tracking method such as ALE (arbitrary Lagrangian-Eulerian) method can be beneficial in the modeling of a drying droplet, whose surface is in continuous deformation \cite{chen2017}. However, in the problem of interest, droplets are steadily fed and preserve their interface shape without requiring a special treatment to track the interface. Therefore, the current study offers a modeling strategy, which handles liquid and gas domains separately but couples them at the interface properly. 

\begin{figure} [h]
\includegraphics[scale=0.9]{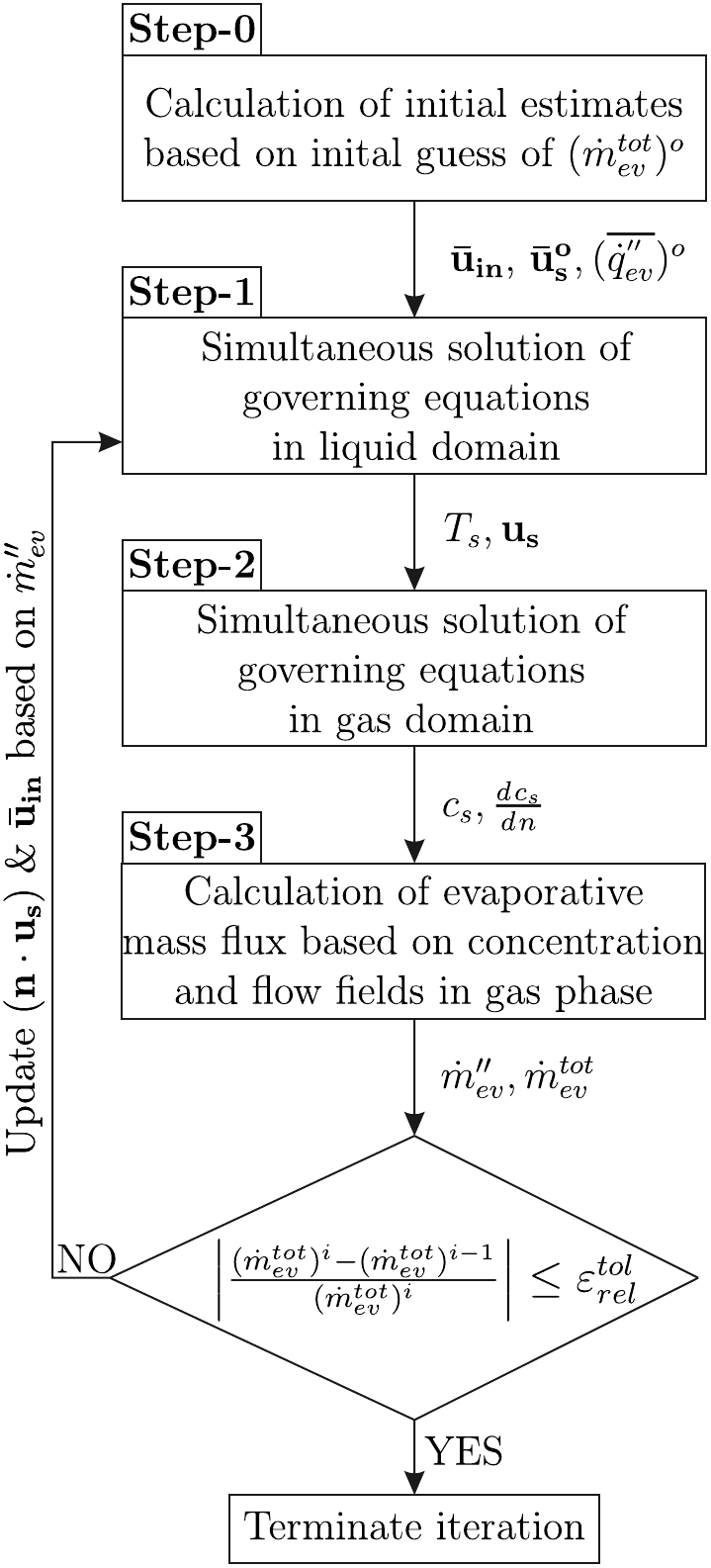}
\centering
\caption{Flowchart of the computational scheme}
\label{fgr:flowchart}
\end{figure}

The proposed model utilizes an iterative computational scheme (\cref{fgr:flowchart}), which enables the simultaneous solution of the steady forms of the mass, momentum, energy, and species conservation equations with temperature dependent thermo-physical properties. Governing equations are solved using the Finite Element Method (FEM) based solver of COMSOL. Linear shape functions are used in the discretization of all variables during FEM formulation. The iterative scheme is implemented using the interface, Livelink\textsuperscript{TM} for MATLAB. Each step of the iterative process is explained in the following sections.

\subsection{Step-0: Initialization}
\label{step-0}
Initial estimates of the boundary conditions at feeding surface ($r=r_i$) and liquid-vapor interface ($r=r_o$) are assigned in the initialization step. A uniform velocity distribution is assumed at the inlet based on the initial guess of total evaporation rate ($\dot{m}^{tot}_{ev}$) from the droplet:

\begin{equation} \label{eqn:u_in}
\bar u_{in}^o=\frac{(\dot{m}^{tot}_{ev})^o}{A_{in} \ \rho} \, ,
\end{equation}

\noindent where $A_{in}$ is the surface area of the inlet surface and the liquid density ($\rho$) is calculated at the substrate temperature ($T_w$). The velocity at the inlet is assumed to be perpendicular to the inlet surface and it is confirmed that direction of the inlet velocity has no observable effect on the results. Distribution of liquid velocity at the liquid-gas interface is assumed uniform in only the first iteration as the initial estimate and calculated based on the inlet-outlet area ratio as follows:

\begin{equation} \label{eqn:u_out0}
\bar u_{s}^o=\bar u_{in} \frac{A_{in}}{A_{out}} \, ,
\end{equation}

\noindent where $A_{out}$ is the area of the droplet surface. Distribution of the interfacial liquid velocity is iterated on within the computational scheme in subsequent iterations. In a similar manner, a homogeneous heat flux is assigned to the interface based on the initial estimate of total evaporation rate:

\begin{equation} \label{eqn:q0}
(\overline{\dot{q}_{ev}^{''}})^o=\frac{(\dot{m}^{tot}_{ev})^o h_{fg}}{A_{out}} \, .
\end{equation}

\noindent This initial estimate is replaced in the following iterations by updating the distribution of evaporative mass flux ($\dot{m}''_{ev}$), which is calculated based on \cref{eqn:evap_flux} after the solution of the gas domain.  

\subsection{Step-1: Solution of the governing equations in liquid domain}
\label{step-1}
Continuity, momentum, and energy equations, Eqs.~(\ref{eqn:vector-cont})~to~(\ref{eqn:vector-en}), with the associated boundary conditions, Eqs.~(\ref{eqn:liq-b1})~to~(\ref{eqn:liq-b4}), are solved simultaneously for the liquid domain using COMSOL which enables the utilization of temperature dependent thermo-physical properties. In this step, solver is applied iteratively to implement both hydrodynamic boundary conditions: normal component of the liquid velocity and tangential stress balance (see Eq.~(\ref{eqn:liq-b4})). Since the solution of the gas domain is not available in Step-1 of the first iteration, temperature field of gas near the interface is unknown. Therefore, interfacial heat conduction to the gas phase is omitted in the first iteration, but included in subsequent iterations.

\subsection{Step-2: Solution of the governing equations in gas domain}
\label{step-2}
Continuity, momentum, energy, and species transport equations, Eqs.~(\ref{eqn:vector-cont})~to~(\ref{eqn:vector-species}), with the associated boundary conditions, Eqs.~(\ref{eqn:gas-b1})~to~(\ref{eqn:gas-b4}), are solved simultaneously for the gas domain with temperature dependent thermo-physical properties. As stated previously, distributions of interfacial gas velocity and temperature are assigned based on the solution of liquid domain and the concept of Stefan flow. Interfacial vapor concentration distribution corresponds to the saturation concentration of vapor at the interfacial temperature, which is implemented by setting the relative humidity to unity in the solver interface.

\subsection{Step-3: Calculation of evaporative mass flux}
\label{step-3}

In Step-3, evaporative mass flux ($\dot{m}''_{ev}(\phi)$) is calculated based on the 
interfacial vapor concentration ($c_s(\phi)$) and its gradient in normal direction ($d \hskip 1 pt c_s(\phi) / {dn}$) as described in \cref{eqn:evap_flux}. Using the evaluated evaporative mass flux and \cref{eqn:mass_balance}, normal component of the interfacial liquid velocity is calculated and stored to be used in the next iterative step. In addition, velocity of the feeding liquid is also updated based on the calculated evaporative mass flux. These updated liquid velocities together with the updated evaporative mass flux are introduced to Step-1 and this iterative procedure is continued until the convergence of total evaporation rate from the droplet.

Embedded grid generator of COMSOL is utilized to mesh the computational domain. The boundary between the liquid and gas domain, \textit{i.e.} the interface, is divided to equal length arcs before meshing. Grid generation starts at the interface on these arcs and continues towards the liquid and gas domains with a certain growth rate. Resolution of the solution is controlled by adjusting the number of boundary elements (\textit{i.e.} arcs), which directly determines the density of the resultant mesh at the interface. Mesh independence test is carried out for all cases simulated by increasing the number of boundary elements, thereby increasing the number of mesh elements. Grid-independence is decided to satisfy when the change in evaporation rate is less than 0.1\%.


\section{Results and Discussion}
\label{sec:results}

The proposed computational framework is tested to model the evaporation from a hemispherical, sessile, and continuously fed water droplet placed on a heated substrate as shown in \cref{fgr:domain}. Two cases with different wall temperatures are simulated. Superheat values---i.e. difference between wall and ambient temperatures---are selected as 9$^{\circ}$C  (\mbox{Case-1}) and 44$^{\circ}$C (\mbox{Case-2}) to demonstrate the effect of increasing substrate temperature. Moreover, both values are  sufficiently high such that utilization of Boussinesq approximation is rendered questionable in the modeling of the buoyancy in the flow solution. The configurations in the simulated two cases match the previous theoretical \cite{akkus2019iterative} and experimental \cite{macdonald2017} studies enabling a direct comparison. The values of the geometric parameters together with far field conditions are provided in \cref{table:props}. The Bond number is calculated as 0.28 confirming the spherical shape of the droplet surface. Emissivity of the water surface is taken as 0.97 \cite{robinson1972}. Binary diffusion coefficient of water vapor in air is estimated based on the temperature dependent formulation suggested in \cite{bolz1976}. For all other thermophysical properties, temperature dependent values are assigned using the material library of COMSOL.

\begin{table}[h]
\caption{Geometrical parameters and far field conditions}
\begin{center}
\begin{tabular}{lll}
\hline 
Feeding opening radius (mm)             & $ r_i $    & 0.175 \\
Droplet radius (mm)             & $ r_o $    & 2.5 \\    
Radius of gas volume (mm)       & $ W $      & 250 \\ 
Height of gas volume (mm)       & $ H $      & 500 \\ 
Ambient pressure (atm)          & $\rm{p_{\infty}}$   & 1 \\ 
Ambient relative humidity       & $ \phi_{RH} $       & 0.25 \\ 
Ambient temperature ($^{\circ}$C)  & $\rm{T_{\infty}}$   & 30 \\ 
Wall temperature in Case-1 ($^{\circ}$C)  & $\rm{T_w}$   & 39 \\ 
Wall temperature in Case-2 ($^{\circ}$C)  & $\rm{T_w}$   & 74 \\  
\hline 
\end{tabular} 
\end{center}
\label{table:props}
\end{table}

The model proposed is also employed to make a benchmark test for the models utilizing simplifying assumptions in droplet evaporation modeling. The comprehensive model (full-model with temperature dependent properties including density) is named as \textbf{FM-1}. Widely utilized Boussinesq approximation (with temperature dependent properties excluding density) is simulated only in gas phase by reflecting the effect of density change on the body force term solely instead of temperature dependent density setting set throughout the domain. This model is called as \textbf{FM-2}. A common approach adopted in the literature is undoubtedly the one accounting for only diffusion of heat and vapor in gas phase because of the its relatively low computational cost; however, it inevitably lacks from the omission of convection transport. To demonstrate the applicability of this approach, simulations are carried out for the diffusion based model (\textbf{DM} hereafter). Moreover, the widely used correlation of Hu and Larson \cite{hu2002evap} is employed to calculate the evaporation rates to exhibit its validity. The last model tested is the one based on a natural convection correlation for a sphere hanging in air environment, which were previously adopted in \cite{ruiz2002,kaneda2008,akkus2017modeling,akkus2019iterative}. Simulations with these models and correlations are performed and evaporation rate predictions of them are summarized in \cref{table:rates}.

Results demonstrate that full-models (\textbf{FM-1} and \textbf{FM-2}) solving the gas flow predict higher evaporation rates compared to those of others. Predictions of \textbf{FM-2} are close to those of \textbf{FM-1} but start to deviate with increasing substrate temperature due to the incapability of Boussinesq approximation in capturing the gas flow field in elevated temperature differentials. Although Stefan flow, thereby the convective mass flux at the interface, is considered, \textbf{DM} severely underestimates evaporation rates due to the absence of convective transport in the gas phase. Correlation proposed by \cite{hu2002evap} is based on the diffusion of vapor and its prediction is expected to be similar to that of \textbf{DM}. However, predictions of them are similar only in the case of Marangoni flow driven droplet. This result is not surprising since the correlation given in \cite{hu2002evap} employs constant interface temperature equal to the substrate temperature and temperature of the droplet surface is closer to the substrate temperature in the presence of Marangoni flow \cite{lu2011,akkus2019iterative}. On the other hand, natural convection correlation based model substantially underestimates the evaporation rate. The poor performance of this model is primarily due to its failure to capture the flow field in the gas. In the problem of interest, buoyancy of gas is triggered by the heated flat substrate, whereas the natural convection correlation is built on a flow field triggered by a sphere hanging in air. Therefore, this model is not applicable in an accurate modeling of the evaporation of a hemispheric droplet since it refers to a geometrically different configuration.


\begin{table}
\centering
\caption{Evaporation rate (in $\mu$g/s) estimation of different models}
\label{table:rates}
\begin{tabular}{lcccc}\\
\hline\\[-3pt]        
 &  \multicolumn{2}{c}{\textbf{w/o Marangoni}}    & \multicolumn{2}{c}{\textbf{w/ Marangoni}}   \\ [3pt]
 & Case-1 & Case-2 & Case-1 & Case-2 \\ [3pt]
\hline\\[-3pt]
Full-model (\textbf{FM-1})  & 19.4  & 135.4 & 24.5 & 238.4  \\[2pt]
Full-model with Boussinesq appr. in gas (\textbf{FM-2})  & 19.4 & 116.2 & 22.9 & 200.0  \\[2pt]
Diffusion (in gas) based model (\textbf{DM}) & 15.0 & 87.2 & 16.8 & 108.6   \\ [3pt]
Diffusion (in gas) based correlation of Hu\&Larson~[33]  & 17.2  & 113.3 & 17.2 & 113.3   \\ [2pt]
Natural conv. (in gas) correlation based model  & 12.1  & 67.8 & 13.4 & 81.3   \\ [2pt]
\hline\\[-2pt]
\end{tabular}
\end{table}

\begin{figure}[h]
\includegraphics[scale=0.65]{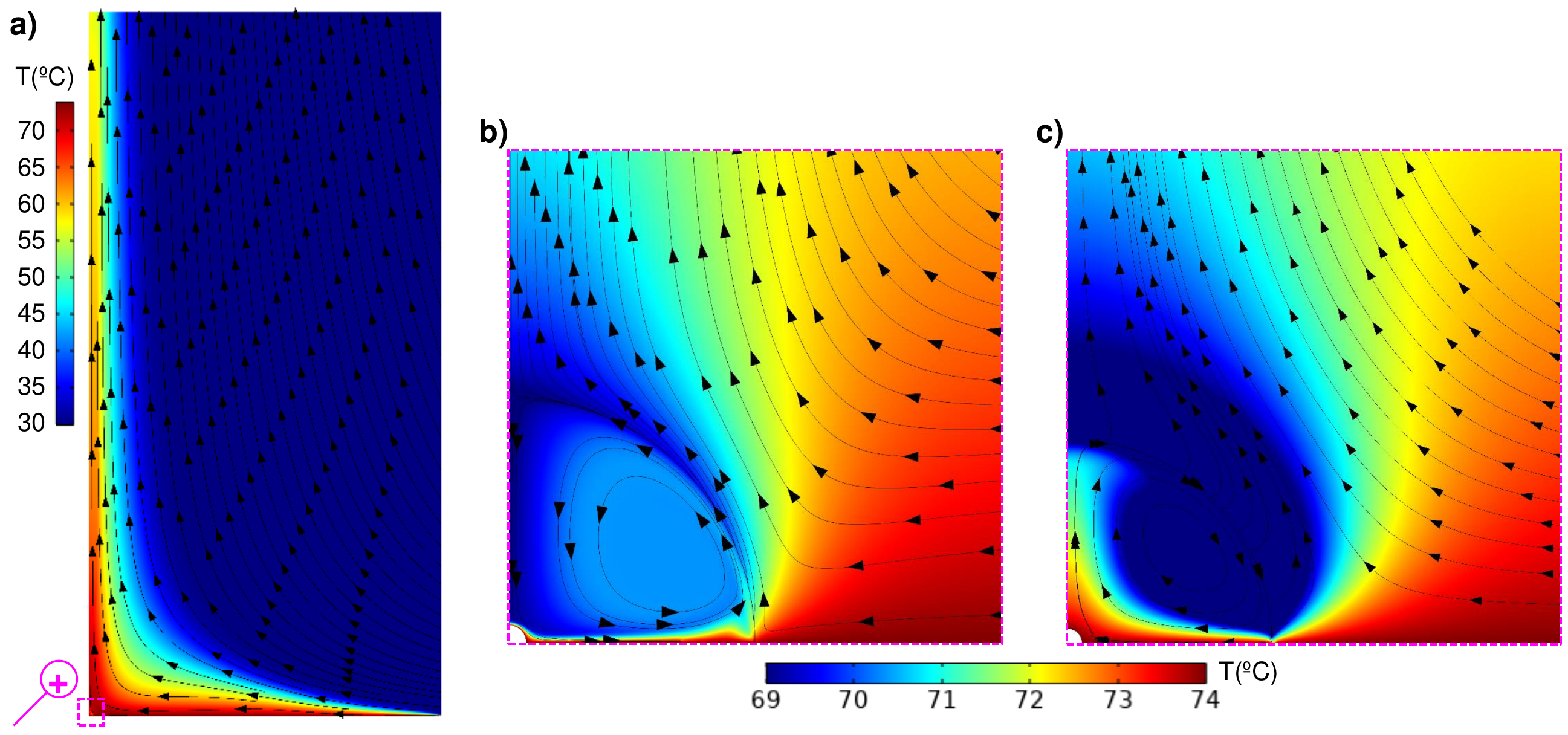}
\centering
\caption{Temperature field and streamlines by \textbf{FM-1} \textbf{a)} in the entire computational domain, \textbf{b)} in droplet and gas region near the droplet surface with thermocapillarity, and \textbf{c)} without thermocapillarity.}
\label{fgr:full_T}
\end{figure}

\begin{figure}[h]
\includegraphics[scale=0.55]{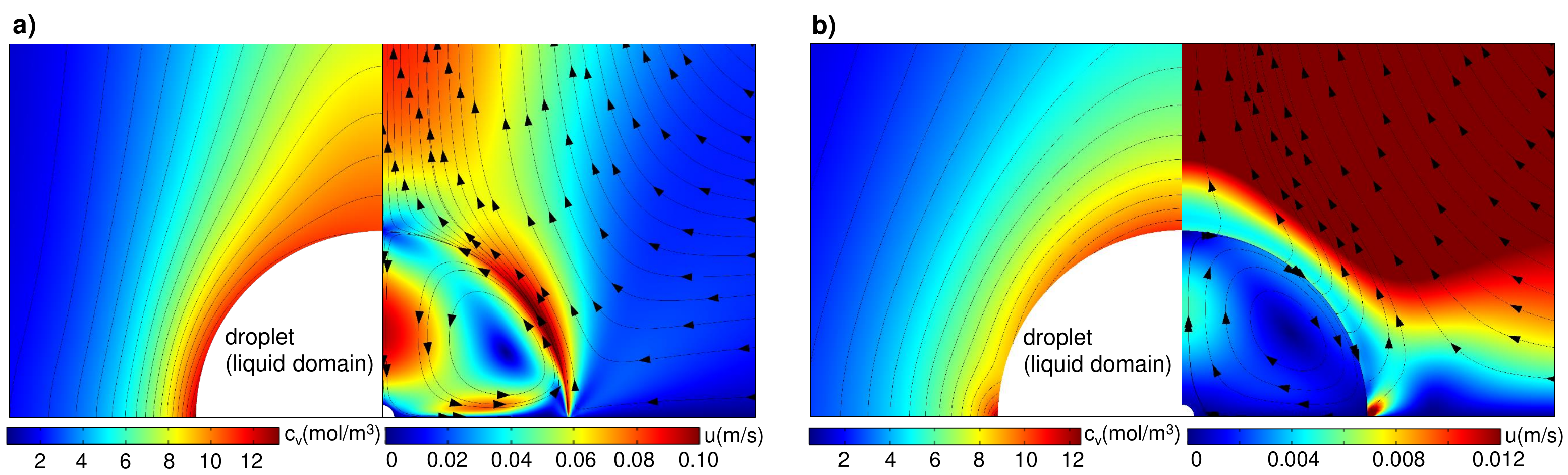}
\centering
\caption{Vapor concentration field (left image) and velocity magnitude field with streamlines (right image) by \textbf{FM-1} \textbf{a)} with and \textbf{b)} without thermocapillarity. Note that scale bars in \textbf{a)} and \textbf{b)} are different for both vapor concentration and velocity magnitude fields.}
\label{fgr:full_u}
\end{figure}

\begin{figure}[h]
\includegraphics[scale=0.75]{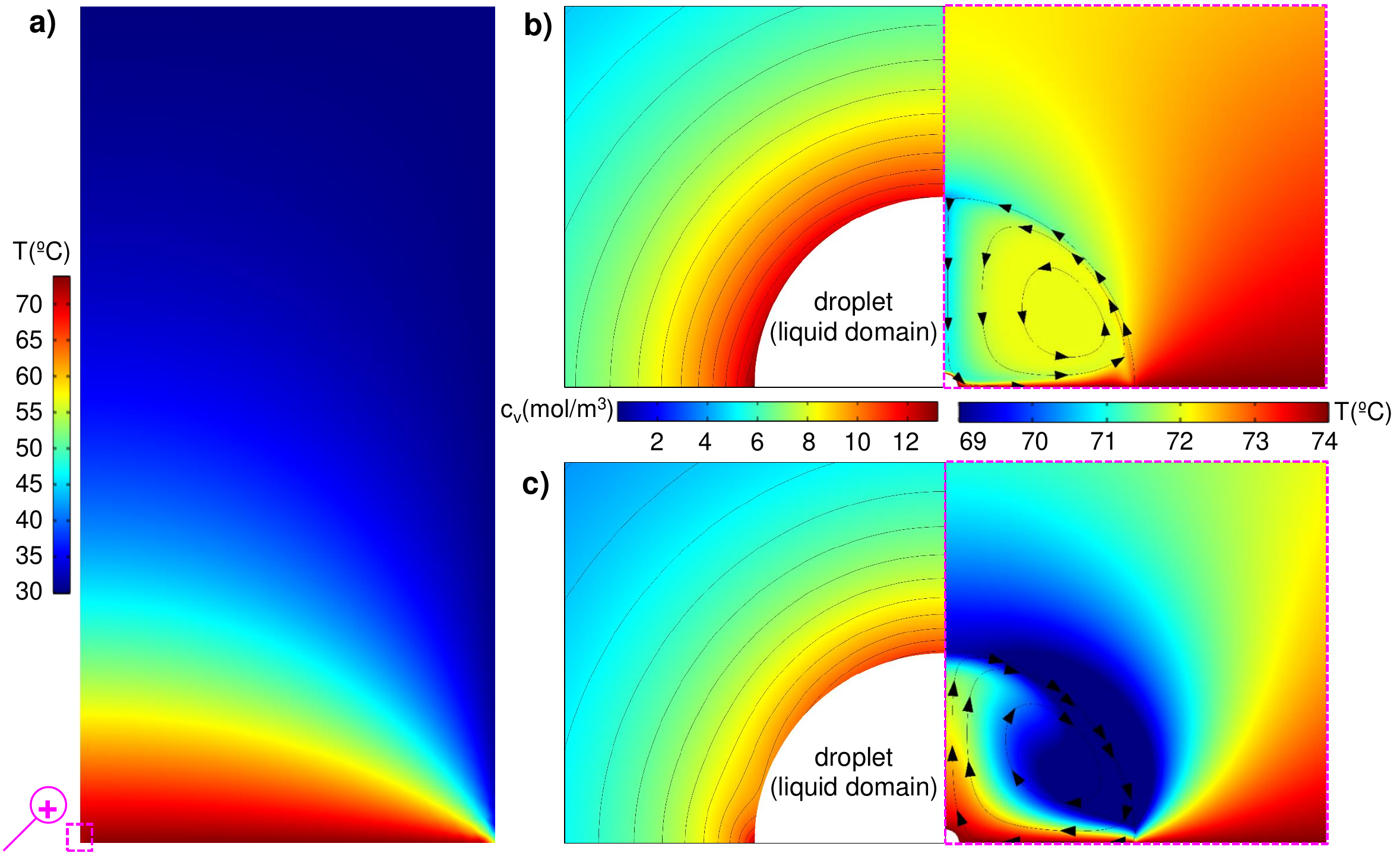}
\centering
\caption{Results of \textbf{DM}. \textbf{a)} Temperature field in the complete domain. Vapor concentration field (left image) and temperature  field with streamlines (right image) \textbf{b)} with and \textbf{c)} without thermocapillarity.}
\label{fgr:diff}
\end{figure}

\begin{figure}[h]
\includegraphics[scale=0.55]{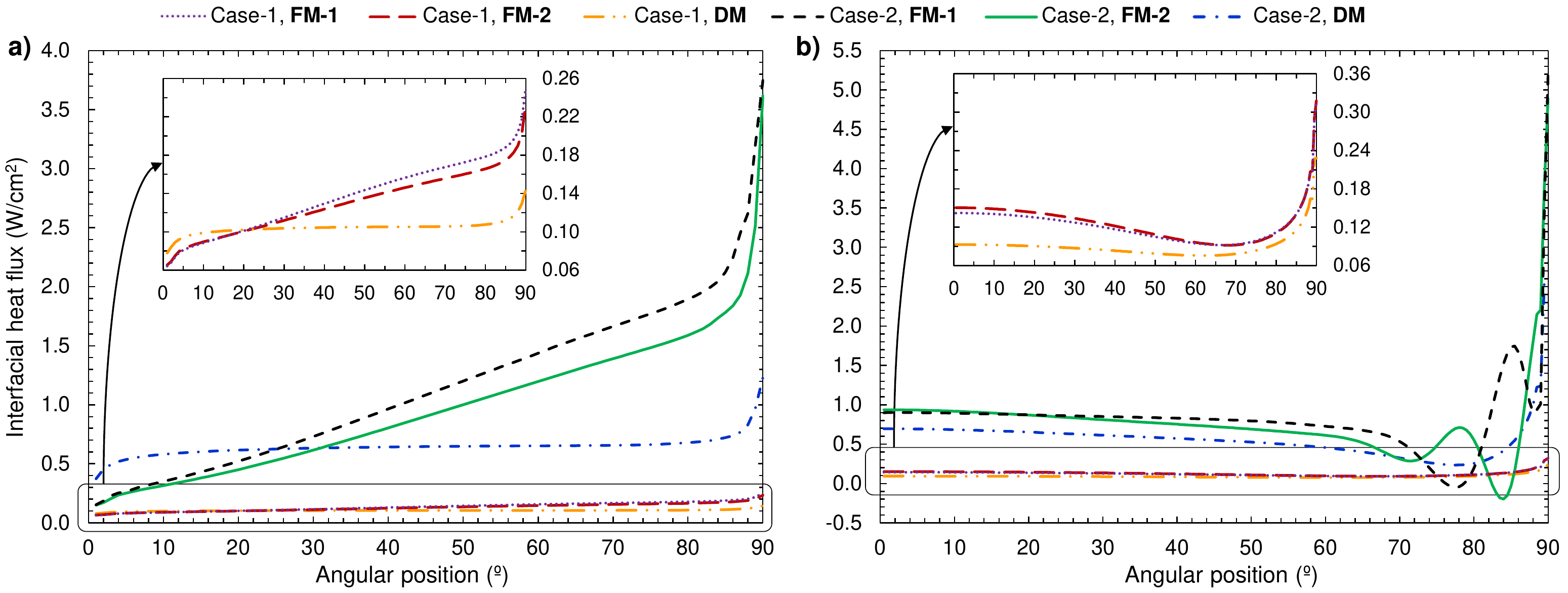}
\centering
\caption{Interfacial heat flux distribution predictions of the models \textbf{a)} with and \textbf{b)} without thermocapillarity.}
\label{fgr:q_dist}
\end{figure}

\begin{figure}[h]
\includegraphics[scale=0.55]{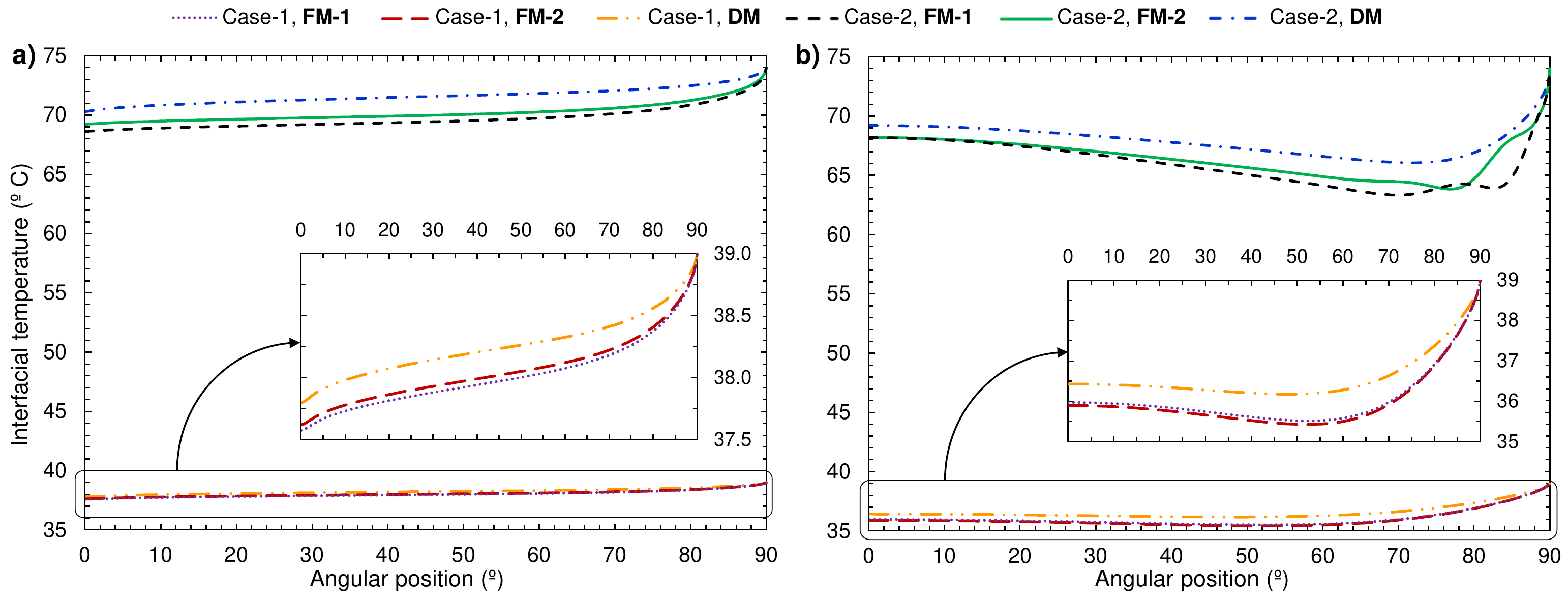}
\centering
\caption{Interfacial temperature distribution predictions of the models \textbf{a)} with and \textbf{b)} without thermocapillarity.}
\label{fgr:T_dist}
\end{figure}

\begin{figure}[h]
\includegraphics[scale=0.52]{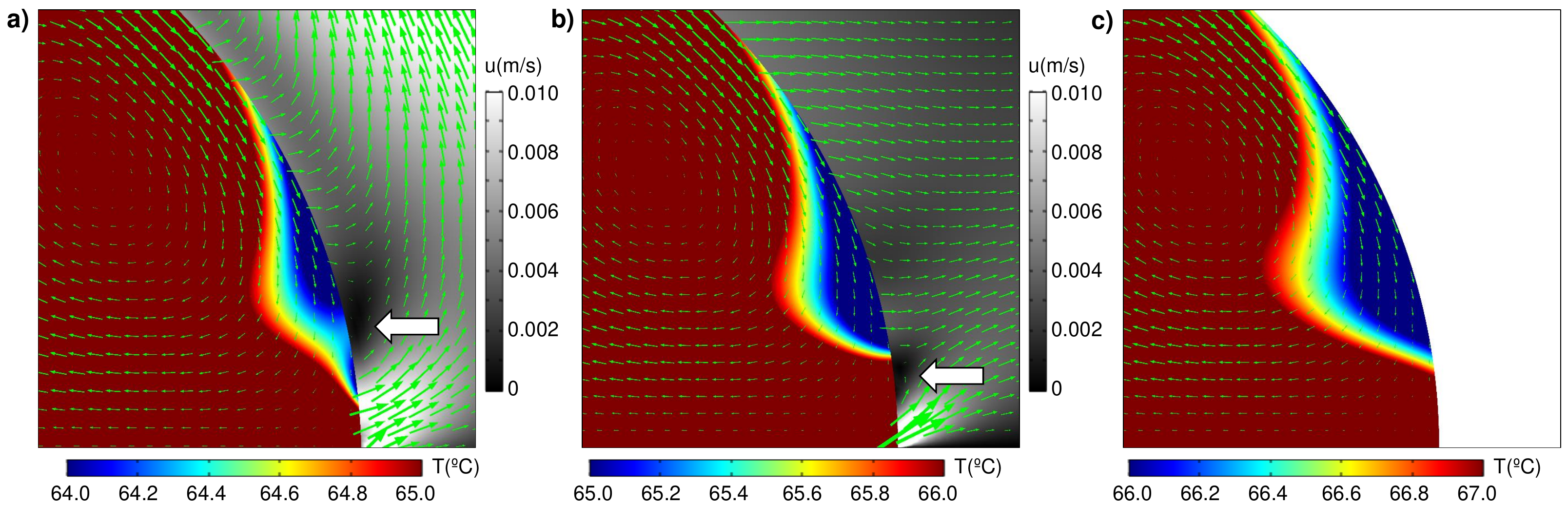}
\centering
\caption{Temperature distribution in liquid and velocity magnitude distribution in gas phase near the contact line with superimposed streamlines as the result of \textbf{a)} \textbf{FM-1}, \textbf{b)} \textbf{FM-2}, and \textbf{c)} \textbf{DM}. White arrows show the zone, where a local vortex leads to small velocity magnitudes. Note that there is no gas flow field in the last plot since \textbf{DM} omits the solution of gas flow.}
\label{fgr:db_dip}
\end{figure}

Although simulations are performed for two cases with different superheats, resultant flow, temperature and concentration fields are reported for only one case (Case-2) in the rest of this study, since the patterns are similar. Figures~\ref{fgr:full_T} and \ref{fgr:full_u} show the temperature and flow fields of \textbf{FM-1}. \Cref{fgr:diff} exhibits the same fields obtained using \textbf{DM}. Whilst \cref{fgr:q_dist} and \ref{fgr:T_dist} show the the distribution of heat flux and temperature along the droplet surface for all models, respectively, \cref{fgr:db_dip} focuses on the temperature patterns near the contact line arising in the absence of Marangoni flow.

A typical natural convection pattern is obtained in the gas domain as a result of \textbf{FM-1} as shown in \cref{fgr:full_T}a, where air enters the domain from the periphery and moves upwards along the axis of the domain due to being heated by the hot wall. This pattern is also quite similar to the results of \textbf{FM-2}. However, not only liquid domain but also gas region near the droplet surface has substantially different resultant patterns based on the presence of thermocapillary flow. When Marangoni effect is included in the model, a strong surface flow in the direction of monotonically decreasing temperature (\textit{i.e.} from the substrate to the apex) results in a main counter clockwise (CCW) vortex (see \cref{fgr:full_T}b) pattern, which is in agreement with previous studies reporting a similar pattern for water drops with contact angles close to 90$^{\circ}$  \cite{ruiz2002,zhang2014temperature,bouchenna2017}. Although the values of Marangoni numbers are much higher than the threshold value \cite{carey} for the cases considered in the current study, the results of the model without Marangoni effect are still reported to demonstrate the underlying physics in the absence of thermocapillary flow, whose presence has been contentious in the literature \cite{ward2004,xu2007} for water droplets and films in air. In \cref{fgr:full_T}c, as opposed to the case with Marangoni flow, a clockwise (CW) vortex pattern appears as a result of the buoyancy of liquid similar to the findings of previous studies \cite{lu2011,akkus2019iterative,pan2020}. It should be noted that buoyancy flow of liquid is dominated by Marangoni flow (\cref{fgr:full_T}b) when thermocapillarity effect is accounted for in accordance with the previous studies \cite{lu2011,bouchenna2017,akkus2019iterative}. Temperature field also substantially differs based on the presence of thermocapillarity. When present, the temperature inside the liquid as well as the gas temperature near the surface increases suggesting a higher convective transport of the energy, which can be understood better by examining the strength of the internal liquid flow.

\Cref{fgr:full_u} shows the distribution of the magnitude of the flow velocity in both liquid and gas domains with and without Marangoni flow together with the distribution of vapor concentration in gas domain, which is presented as the mirror image at left sides of the figures. Results clearly exhibit that Marangoni flow is much stronger than buoyancy flow explaining the elevated amount of heat transport (convective) from the substrate to the interface. Gas flow near the interface is also affected by the interface velocity. When thermocapillary flow is effective, gas flow is in the same direction with the interface velocity, which results in the upward acceleration of gas along the interface (\cref{fgr:full_u}a). On the contrary, when thermocapillary flow is absent, buoyancy driven surface flow (from apex to the wall) is against the direction of gas flow, which decelerates the gas near the interface leading to the bending of the streamlines as shown in \cref{fgr:full_u}b. Moreover, in the case of buoyancy driven internal liquid flow, Stefan flow of the gas mixture originating from the drop surface is apparent. This flow intensifies near the contact line due to the substantially increased rate of evaporation by manifesting a velocity jet seen in \cref{fgr:full_u}b similar to the finding of \cite{pan2020}. Distributions of vapor concentration in \cref{fgr:full_u} are consistent with the flow fields. In the case of Marangoni flow, tangential upward movement of the gas flow carries the vapor to the apex region, where concentration isolines are distorted in the upward direction due to the accumulation of vapor. This accumulation is less apparent in the absence of thermocapillary flow since Stefan flow aids transportation of vapor in the radial direction. In addition, the elevated evaporation rate results in the denser vapor zone near the contact line as seen in \cref{fgr:full_u}b.    

Although gas flow has a dominant effect on evaporation rate, diffusion based models are frequently applied for the estimation of evaporation rate. When the gas flow is not taken into account, conduction heat transfer becomes effective in energy transport resulting in a stratified temperature distribution in the gas domain as seen in \cref{fgr:diff}a. 
\noindent Due to the decrease in evaporation rate, evaporative cooling is also reduced, which leads to warmer drops (\cref{fgr:diff}b~and~c) than the ones obtained in the presence of gas flow (\cref{fgr:full_T}b~and~c). Although the flow patterns of liquid inside the droplet are unaffected, distribution of vapor concentration inevitably differs in the absence gas flow. This difference is especially apparent in the case with Marangoni convection. Instead of elongated isolines, stratified distribution of vapor concentration shows the lack of convective vapor transport in \cref{fgr:diff}b. Likewise, in the case of buoyancy driven liquid flow, less distorted concentration isolines form as seen in \cref{fgr:diff}c. However, non-uniformity of vapor concentration near the interface is still present, which can be linked to the non-uniformity of evaporation distribution along the interface. A better understanding requires the examination of heat flux distribution along the interface.     

Regardless of the presence of thermocapillarity, substrate temperature or the model utilized, interfacial heat flux reaches a maximium at the contact line as seen in \cref{fgr:q_dist}. 
\noindent However, the increase of heat flux towards the contact line is monotonic in the presence of Marangoni flow (\cref{fgr:q_dist}a), whereas it is nonmonotanic in the case of buoyancy driven internal liquid flow (see \cref{fgr:q_dist}b). Although both \textbf{FM-1} and \textbf{FM-2} exhibit a near linear variation (except $\sim$10$^{\circ}$ portion adjacent the contact line, where flux values increase rapidly) in the case of Marangoni flow, \textbf{FM-1} predicts higher heat flux along the interface than \textbf{FM-2}. \textbf{DM} predicts near uniform (except for $\sim$5$^{\circ}$ portions adjacent to the apex and contact line) interfacial flux and it underestimates the heat flux along the interface (except $\sim$25$^{\circ}$ portion adjacent to the apex) with respect to \textbf{FM-1} when Marangoni flow is present. This underestimation increases towards the contact line, which is in accordance with the literature \cite{saada2010}. While Case-1 yields lower interfacial heat flux values as expected (the inset of \cref{fgr:q_dist}a), both cases result in similar distribution patterns along the interface. 

In the case of buoyancy driven liquid flow, all models have a slightly decreasing heat flux distribution starting from the apex. However, substrate temperature greatly affects the trend of the distributions approaching the contact line. Case-1 yields a minimum flux point around the angular position of 70$^{\circ}$ in all models. While \textbf{DM} underestimates the flux values throughout the surface with respect to full-models (\textbf{FM-1} and \textbf{FM-2}), resultant flux distribution of full models are nearly identical. In Case-2, \textbf{DM} yields a similar flux pattern to the one in \mbox{Case-1}. However, full models exhibit two dips and one peak in between the dips before the final rise at the contact line, a behavior requiring a close examination. Another conspicuous result is the prediction of negative flux values by full-models. Negative flux means heat transfer from ambient to the droplet. Although perplexing, this result is understandable since the surrounding bulk gas flow is warmer than the ambient due to the natural convection pattern shown in \cref{fgr:full_T}a, which in turn, leads to a net conduction heat transfer from the gas to the interface. Whenever this conduction heat transfer becomes larger than the total of evaporative and radiative heat transfer, net interfacial flux changes its sign.

In the presence of Marangoni flow, temperature distribution predictions of all models for both cases are similar: monotonically increasing temperature towards the contact line as shown in \cref{fgr:T_dist}. 
\noindent Since the resultant evaporation rate of \textbf{FM-1} is higher than that of the others, this model predicts the lowest apex temperature. In the case of buoyancy driven liquid flow, a temperature dip appears in both substrate temperatures but it is more apparent for higher substrate temperature. Presence of temperature dip was previously reported in experimental \cite{macdonald2017}  and numerical  \cite{akkus2019iterative} studies. However, in the current study, \textbf{FM-1} predicts two temperature dips, which, to the best of the authors' knowledge, has never been reported in the literature. Reason of the formation of two dips is due to the formation of a local temperature peak in the region of minimum temperature and this local peak can be attributed to the heating effect due to conduction heat transfer from the gas. However, local temperature peak does not appear as a result of t\textbf{FM-2}. In order to elucidate this, the local liquid temperature distribution together with the local velocity magnitude distribution in the gas phase are plotted in \cref{fgr:db_dip}.

In the buoyancy driven liquid flow, internal convection pattern drives the warm liquid from substrate to the apex, then it moves towards the contact line along the interface, during which the liquid cools down due to the evaporative heat loss. However, conduction heat transfer from hot substrate starts to heat the liquid near the contact line, which yields an intermediate cooler zone above the substrate (these temperature dips are made more apparent by adjusting the color ranges in \cref{fgr:db_dip}). However, as shown in \cref{fgr:db_dip}a, \textbf{FM-1} exhibits a secondary temperature dip between the primary one and the substrate due to the local heating, which arises because of the suppression of evaporation. White arrow in \cref{fgr:db_dip}a indicates the region with substantially low velocities. This region appears due to the presence of a local vortex which forms under the effect of strong Stefan flow originating from the contact line region. Local vortex suppresses the convective transport of vapor from the interface, which reduces the evaporation rate, and consequently, the evaporative cooling. Therefore, the liquid is subjected to an immediate heating. After passing the vortex zone, strong Stefan flow suddenly enhances the evaporation resulting in the secondary temperature dip. Following the dip, conduction from the hot substrate raises the interfacial temperature. Secondary dip is not shown in the results of \textbf{FM-2} (see \cref{fgr:db_dip}b) due to the location of vortex region, which is closer to the contact line. Despite the elevated evaporative cooling beyond the vortex zone, heat transfer from the substrate heats up the liquid near the contact line. Consequently, a secondary dip cannot form. Different vortex location predictions of full-models are linked to the different flow fields in the gas phase. While \textbf{FM-2} predicts a gas flow mainly in longitudinal direction, \textbf{FM-1} is able to capture the rise of the gas immediately, which pushes the vortex zone above compared to \textbf{FM-2}. \textbf{DM}, on the other hand, always predicts a single temperature well when the internal liquid flow is buoyancy driven.
 
\section{Summary and Conclusions}
\label{sec:conclusion}

A theoretical framework is introduced to model the steady droplet evaporation with all relevant physics and a test case---evaporation from a continuously fed constant shape hemispherical water droplet resting on a heated flat substrate---is solved. Computational performance of the proposed model, which utilizes temperature dependent thermophysical properties, is compared with several simplified models adopting widely used assumptions in the literature. Simulations demonstrate that the comprehensive model introduced in the current study (\textbf{FM-1}) yields the highest evaporation rates compared to other simplified models. When the effect of gas convection is excluded by employing a diffusion based model in the gas phase, evaporation rates are underestimated by 23--54\% with respect to \textbf{FM-1}. Implementation of Boussinesq approximation in the gas phase yields the underestimation of evaporation by 14--16\% for the high substrate temperature cases (superheat value of 44 K), while underestimation is less than 6\% in low substrate temperature cases (superheat value of 9 K). Deviations of simplified models increase not only with increasing substrate temperature but also with the presence of Marangoni flow (up to 50\% increase with respect to the case with buoyancy driven liquid flow for the high substrate temperature). In the absence of Marangoni flow, all models predict a temperature well zone at the interface but the comprehensive model identifies a secondary dip in this zone, which is reported for the first time in the literature. Based on these findings conclusions of the present study are:

\begin{itemize}

\item Boussinesq approximation fails to capture the physics in the gas phase for the cases with elevated substrate temperatures, which manifests the need of using temperature dependent thermophysical properties in such cases.

\item Diffusion limited approaches in the modeling of gas domain are inadequate in simulating the physics since they lead to substantial discrepancies in the prediction of evaporation rates. 

\item Utilization of natural convection or diffusion based correlations in the estimation the interfacial evaporation flux without solving the gas phase is not appropriate since the evaporation flux is highly sensitive to the physics of the gas flow near the interface, which is specific to the instantaneous configuration of the problem such as contact angle of the droplet, substrate temperature or the concentration of vapor in ambient air. 

\item Physics of the fluid in one phase inevitably affects the other phase as long as proper coupling is established at the interface. For instance, convection mechanism inside the droplet (buoyancy or Marangoni driven) regulates the interfacial velocities, which determines the flow field of gas near the interface. Likewise, natural convection pattern of gas has paramount effect on the evaporation rate of the droplet. Therefore, omission of a physical phenomenon in one of the two phases cannot be justified while being considered in the other phase.    

\end{itemize}

Theoretical framework established in this study is demonstrated to be effective in creating models of steady droplet evaporation, which are able to capture unreported physics of droplet evaporation. Using this framework, our future work will focus on studying the evaporation from droplets with various contact angles and sizes in different environmental and substrate configurations with the ultimate objective of maximizing cooling rates in droplet based thermal management applications.

\section*{Acknowledgments}

B.\c{C}. would like to acknowledge fundings from the Turkish Academy of Sciences through Outstanding Young Scientist Program (T\"UBA-GEB\.IP) and The Science Academy, Turkey through Turkey Distinguished Young Scientist Award (BAGEP).

\addcontentsline{toc}{section}{References}

\bibliographystyle{unsrt}

\bibliography{references}

\begin{thebibliography}{10}

\bibitem{smalyukh2006}
I.~I. Smalyukh, O.~V. Zribi, J.~C. Butler, O.~D. Lavrentovich, and G.~C.~L.
  Wong.
\newblock Structure and dynamics of liquid crystalline pattern formation in
  drying droplets of dna.
\newblock {\em Phys. Rev. Lett.}, 96(17):177801, 2006.

\bibitem{lim2009evaporation}
T.~Lim, J.~Jeong, J.~Chung, and J.~T. Chung.
\newblock Evaporation of inkjet printed pico-liter droplet on heated substrates
  with different thermal conductivity.
\newblock {\em J. Mech. Sci. Technol.}, 23(7):1788--1794, 2009.

\bibitem{ebrahimi2013}
A.~Ebrahimi, P.~Dak, E.~Salm, S.~Dash, S.~V. Garimella, R.~Bashir, and M.~A.
  Alam.
\newblock Nanotextured superhydrophobic electrodes enable detection of
  attomolar-scale dna concentration within a droplet by non-faradaic impedance
  spectroscopy.
\newblock {\em Lab Chip}, 13(21):4248--4256, 2013.

\bibitem{wu2014}
H.~Wu, L.X. Chen, X.Q. Zeng, T.H. Ren, and W.~H. Briscoe.
\newblock Self-assembly in an evaporating nanofluid droplet: rapid
  transformation of nanorods into 3d fibre network structures.
\newblock {\em Soft Matter}, 10(29):5243--5248, 2014.

\bibitem{bar2006}
A.~Bar-Cohen, M.~Arik, and M.~Ohadi.
\newblock Direct liquid cooling of high flux micro and nano electronic
  components.
\newblock {\em Proceedings of the IEEE}, 94(8):1549--1570, 2006.

\bibitem{liang2017}
G.~Liang and I.~Mudawar.
\newblock Review of spray cooling--part 1: Single-phase and nucleate boiling
  regimes, and critical heat flux.
\newblock {\em Int. J. Heat Mass Tran.}, 115:1174--1205, 2017.

\bibitem{kokalj2010}
T.~Kokalj, H.~Cho, M.~Jenko, and L.~P. Lee.
\newblock Biologically inspired porous cooling membrane using arrayed-droplets
  evaporation.
\newblock {\em Appl. Phys. Lett.}, 96(16):163703, 2010.

\bibitem{macdonald2017b}
S.~Chakraborty, Rosen.~M. A., and B.~D. MacDonald.
\newblock Analysis and feasibility of an evaporative cooling system with
  diffusion-based sessile droplet evaporation for cooling microprocessors.
\newblock {\em Appl. Therm. Eng.}, 125:104--110, 2017.

\bibitem{duh1989}
J.~C. Duh and Wen-Jei Yang.
\newblock Numerical analysis of natural convection in liquid droplets by phase
  change.
\newblock {\em Numerical Heat Transfer}, 16(2):129--154, 1989.

\bibitem{lozinski1993}
D.~Lozinski and M.~Matalon.
\newblock Thermocapillary motion in a spinning vaporizing droplet.
\newblock {\em Phys. Fluids A-Fluid}, 5(7):1596--1601, 1993.

\bibitem{shih1996}
A.~T. Shih and C.~M. Megaridis.
\newblock Thermocapillary flow effects on convective droplet evaporation.
\newblock {\em Int. J. Heat Mass Tran.}, 39(2):247--257, 1996.

\bibitem{ruiz2002}
O~.E. Ruiz and W.~Z. Black.
\newblock Evaporation of water droplets placed on a heated horizontal surface.
\newblock {\em J. Heat Transf.}, 124, 2002.

\bibitem{girard2006}
F.~Girard, M.~Antoni, S.~Faure, and A.~Steinchen.
\newblock Evaporation and marangoni driven convection in small heated water
  droplets.
\newblock {\em Langmuir}, 22(26):11085--11091, 2006.

\bibitem{hu2005}
H.~Hu and R.G. Larson.
\newblock Analysis of the effects of marangoni stresses on the microflow in an
  evaporating sessile droplet.
\newblock {\em Langmuir}, 21(9):3972--3980, 2005.

\bibitem{xu2007}
X.~Xu and J.~Luo.
\newblock Marangoni flow in an evaporating water droplet.
\newblock {\em Appl. Phys. Lett.}, 91(12):124102, 2007.

\bibitem{ristenpart2007}
W.~D. Ristenpart, P.~G. Kim, C.~Domingues, J.~Wan, and H.~A. Stone.
\newblock Influence of substrate conductivity on circulation reversal in
  evaporating drops.
\newblock {\em Phys. Rev. Lett.}, 99(23):234502, 2007.

\bibitem{kaneda2008}
M.~Kaneda, K.~Hyakuta, Y.~Takao, H.~Ishizuka, and J.~Fukai.
\newblock Internal flow in polymer solution droplets deposited on a lyophobic
  surface during a receding process.
\newblock {\em Langmuir}, 24(16):9102--9109, 2008.

\bibitem{xu2009}
X.~Xu, J.~Luo, and D.~Guo.
\newblock Criterion for reversal of thermal marangoni flow in drying drops.
\newblock {\em Langmuir}, 26(3):1918--1922, 2009.

\bibitem{yoshitake2010}
Y.~Yoshitake, S.~Yasumatsu, M.~Kaneda, K.~Nakaso, and J.~Fukai.
\newblock Structure of circulation flows in polymer solution droplets receding
  on flat surfaces.
\newblock {\em Langmuir}, 26(6):3923--3928, 2010.

\bibitem{lu2011}
G.~Lu, Y.~Y. Duan, X.~D. Wang, and D.~J. Lee.
\newblock Internal flow in evaporating droplet on heated solid surface.
\newblock {\em Int. J. Heat Mass Tran.}, 54(19):4437--4447, 2011.

\bibitem{zhang2014temperature}
K.~Zhang, L.~Ma, X.~Xu, J.~Luo, and D.~Guo.
\newblock Temperature distribution along the surface of evaporating droplets.
\newblock {\em Phys. Rev. E}, 89(3):032404, 2014.

\bibitem{barash2015}
L.~Y. Barash.
\newblock Dependence of fluid flows in an evaporating sessile droplet on the
  characteristics of the substrate.
\newblock {\em Int. J. Heat Mass Tran.}, 84:419--426, 2015.

\bibitem{bouchenna2017}
C.~Bouchenna, M.~A. Saada, S.~Chikh, and L.~Tadrist.
\newblock Generalized formulation for evaporation rate and flow pattern
  prediction inside an evaporating pinned sessile drop.
\newblock {\em Int. J. Heat Mass Tran.}, 109:482--500, 2017.

\bibitem{josyula2018}
T.~Josyula, Z.~Wang, A.~Askounis, D.~Orejon, S.~Harish, Y.~Takata, P.~S.
  Mahapatra, and A.~Pattamatta.
\newblock Evaporation kinetics of pure water drops: Thermal patterns, marangoni
  flow, and interfacial temperature difference.
\newblock {\em Phys. Rev. E}, 98(5):052804, 2018.

\bibitem{wang2018}
L.~Wang and M.~T. Harris.
\newblock Stagnation point of surface flow during drop evaporation.
\newblock {\em Langmuir}, 34(20):5918--5925, 2018.

\bibitem{akkus2019iterative}
Y.~Akkus, B.~{\c{C}}etin, and Z.~Dursunkaya.
\newblock An iterative solution approach to coupled heat and mass transfer in a
  steadily fed evaporating water droplet.
\newblock {\em J. Heat Transf.}, 141(3), 2019.

\bibitem{kelly2011}
P.~L. Kelly-Zion, C.~J. Pursell, S.~Vaidya, and J.~Batra.
\newblock Evaporation of sessile drops under combined diffusion and natural
  convection.
\newblock {\em Colloid. Surface. A}, 381(1-3):31--36, 2011.

\bibitem{sobac2012pre}
B.~Sobac and D.~Brutin.
\newblock Thermal effects of the substrate on water droplet evaporation.
\newblock {\em Phys. Rev. E}, 86(2):021602, 2012.

\bibitem{carle2013}
F.~Carle, B.~Sobac, and D.~Brutin.
\newblock Experimental evidence of the atmospheric convective transport
  contribution to sessile droplet evaporation.
\newblock {\em Appl. Phys. Lett.}, 102(6):061603, 2013.

\bibitem{carle2016}
F.~Carle, S.~Semenov, M.~Medale, and D.~Brutin.
\newblock Contribution of convective transport to evaporation of sessile
  droplets: empirical model.
\newblock {\em Int. J. Therm. Sci.}, 101:35--47, 2016.

\bibitem{saada2010}
M.~A. Saada, S.~Chikh, and L.~Tadrist.
\newblock Numerical investigation of heat and mass transfer of an evaporating
  sessile drop on a horizontal surface.
\newblock {\em Phys. Fluids}, 22(11):112115, 2010.

\bibitem{chen2017}
Y.~H. Chen, W.~N. Hu, J.~Wang, F.~J. Hong, and P.~Cheng.
\newblock Transient effects and mass convection in sessile droplet evaporation:
  The role of liquid and substrate thermophysical properties.
\newblock {\em Int. J. Heat Mass Tran.}, 108:2072--2087, 2017.

\bibitem{hu2002evap}
H.~Hu and R.~G. Larson.
\newblock Evaporation of a sessile droplet on a substrate.
\newblock {\em J. Phys. Chem. B}, 106(6):1334--1344, 2002.

\bibitem{deegan1997}
R.~D. Deegan, O.~Bakajin, T.~F. Dupont, G.~Huber, S.~R. Nagel, and T.~A.
  Witten.
\newblock Capillary flow as the cause of ring stains from dried liquid drops.
\newblock {\em Nature}, 389(6653):827--829, 1997.

\bibitem{deegan2000}
R.~D. Deegan, O.~Bakajin, T.~F. Dupont, G.~Huber, S.~R. Nagel, and T.~A.
  Witten.
\newblock Contact line deposits in an evaporating drop.
\newblock {\em Phys. Rev. E}, 62(1):756, 2000.

\bibitem{popov2005}
Y.~O. Popov.
\newblock Evaporative deposition patterns: spatial dimensions of the deposit.
\newblock {\em Phys. Rev. E}, 71(3):036313, 2005.

\bibitem{pan2020}
Z.~Pan, J.~A. Weibel, and S.~V. Garimella.
\newblock Transport mechanisms during water droplet evaporation on heated
  substrates of different wettability.
\newblock {\em Int. J. Heat Mass Tran.}, 152:119524, 2020.

\bibitem{girard2008}
F.~Girard, M.~Antoni, and K.~Sefiane.
\newblock On the effect of marangoni flow on evaporation rates of heated water
  drops.
\newblock {\em Langmuir}, 24(17):9207--9210, 2008.

\bibitem{akkus2017modeling}
Y.~Akku{\c{s}}, B.~{\c{C}}etin, and Z.~Dursunkaya.
\newblock Modeling of evaporation from a sessile constant shape droplet.
\newblock In {\em ASME 15th International Conference on Nanochannels,
  Microchannels, and Minichannels}, page V001T04A004. American Society of
  Mechanical Engineers, 2017.

\bibitem{strotos2008}
G.~Strotos, M.~Gavaises, A.~Theodorakakos, and G.~Bergeles.
\newblock Numerical investigation on the evaporation of droplets depositing on
  heated surfaces at low weber numbers.
\newblock {\em Int. J. Heat Mass Tran.}, 51(7):1516--1529, 2008.

\bibitem{macdonald2017}
M.~A. Mahmud and B.~D. MacDonald.
\newblock Experimental investigation of interfacial energy transport in an
  evaporating sessile droplet for evaporative cooling applications.
\newblock {\em Phys. Rev. E}, 95, 2017.

\bibitem{Ferziger02}
Johel~H. Ferziger and Milovan Peri\'c.
\newblock {\em Computational Methods for Fluid Dynamics}.
\newblock Springer, New York, 3$^{rd}$ edition, 2002.

\bibitem{prosperetti1979}
A.~Prosperetti.
\newblock Boundary conditions at a liquid-vapor interface.
\newblock {\em Meccanica}, 14(1):34--47, 1979.

\bibitem{robinson1972}
P.~J. Robinson and J.~A. Davies.
\newblock Laboratory determinations of water surface emissivity.
\newblock {\em J. Appl. Meteor.}, 11(8):1391--1393, 1972.

\bibitem{bolz1976}
R.~Bolz and G.~Tuve.
\newblock {\em Handbook of Tables for Applied Engineering Science}.
\newblock CRC Press, Cleveland, 2$^{nd}$ edition, 1976.

\bibitem{carey}
V.~P. Carey.
\newblock {\em Liquid-vapor Phase Change Phenomena}.
\newblock Hemisphere Publishing House, New York, 1992.

\bibitem{ward2004}
C.~A. Ward and F.~Duan.
\newblock Turbulent transition of thermocapillary flow induced by water
  evaporation.
\newblock {\em Phys. Rev. E}, 69(5):056308, 2004.

\end{thebibliography}

\section*{Declarations of interest}
\addcontentsline{toc}{section}{Competing_interests}
None.

\end{document}